\documentclass[aps,prl,twocolumn,nofootinbib]{revtex4-2}
\usepackage{amsmath,amssymb,bm}
\usepackage{mathtools}
\usepackage{physics}
\usepackage{hyperref}

\begin{document}

\title{Global Integrated Null Energy Contribution : Classification of Traversable Wormholes}

\author{Soumya Chakrabarti}
\email{soumya.chakrabarti@vit.ac.in}
\affiliation{School of Advanced Sciences, Vellore Institute of Technology,
Tiruvalam Rd, Katpadi, Vellore, Tamil Nadu 632014 India}

\begin{abstract}
We argue that the local violation of null energy condition at the throat of a traversable wormhole does not necessarily indicate that the total volume integrated null energy contribution is negative. It is already known that it can be arbitrarily close to zero. We show that it can even be positive and use this argument to reconstruct a family of wormhole geometries. We also propose a novel classification of traversable wormholes based on the global volume integrated measure of null energy condition.
\end{abstract}

\maketitle

\paragraph*{\bf Introduction}
Traversable wormholes are always associated with a \textit{throat}, a tunnel where a family of converging radial null rays become parallel. The throat imposes a local geometric requirement called the \textit{flare-out} condition which necessarily leads to a local violation of the null energy condition (NEC) \cite{mt1, mt2}. In extension, this can be associated with an averaged null energy condition (ANEC) of the entire geometry, provided the violation is considered along suitable null geodesics \cite{friedman, gall, hochberg1, hochberg2}. This can raise some questions related to the strength of the violation; for instance, is a substantial amount of NEC-violation required to sustain a traversable wormhole, or can the required violation be made arbitrarily small? This question was answered Visser, Kar and Dadhich, who introduced a suitable volume-integral measure to derive the total exoticity of a wormhole geometry. They proved that traversable wormholes can be supported by a matter distribution which violates NEC locally and the ANEC along a null geodesic, but the total volume integrated measure of NEC violation can be made arbitrarily small \cite{KDV2003}. Motivated by the measure of a global violation of the energy condition we extend this question : does a violation of local energy condition enforced by a wormhole throat also enforce a global violation of the integrated energy condition? In simpler words, must a wormhole that violates the NEC at its throat also have a negative volume-integrated NEC? \medskip

In this Letter, we show that the answer is \textit{`Not Necessarily'}. We demonstrate that there exists a family of traversable wormholes for which the global volume-integrated NEC contribution is positive, while the local geometry continues to satisfy the near-throat violation of NEC as well as ANEC. For a static, spherically symmetric wormhole metric we define the cumulative form of the volume integrated measure of NEC as
\begin{equation}\label{gamma1}
\Gamma(r) = -\int_{r_0}^{r}\left[1-B'(x)\right]\ln\!\left[ \frac{e^{2\Phi(x)}}{1-B(x)/x} \right]dx.
\end{equation}
$B(r)$ is the shape function and $\Phi(r)$ is the redshift function, respectively. The asymptotic value of the integral, i.e., of $\Gamma$, gives the corresponding volume-integrated global NEC contribution. Using $\Gamma$, we classify families of wormhole solutions based on their volume-integrated radial NEC contribution. Moreover, we show that once a desired globally integrated NEC profile is specified through $\Gamma(r)$, there can always be a reconstruction of the redshift function, keeping the local NEC violation preserved. \medskip

This question is particularly relevant in view of the role played by energy conditions in General Relativity \cite{epstein, hochberg3, ford1, borde, ford2}. The area-increase theorem, for example, is overturned by the cumulative null-energy violations induced by quantum effects, as evident in Hawking radiation \cite{hawk, bir}. Similarly, the singularity theorems rely on the energy conditions to convert the local focusing of geodesics into global concepts of geodesic incompleteness \cite{schon1, schon2, barcelo}. Traversable wormholes provide yet another setting in which the interplay between local energy conditions and global geometry becomes essential. The flare-out condition requires NEC violation at the throat, but it remains unclear whether this local requirement necessarily fixes the total integrated energy budget of the entire geometry. Can the cumulative null energy contribution be redistributed away from the wormhole throat while preserving the required local exoticity? It is this distinction between a \textit{local necessity} and the \textit{global consequence} that motivates our analysis.   \medskip

\paragraph*{\bf Total Exoticity : A Volume Integral Measure of Null Energy} 
For a static, spherically symmetric spacetime written in Morris-Thorne form \cite{mt1, mt2}
\begin{equation}
ds^2 = -e^{2\Phi(r)}dt^2 + \frac{dr^2}{1-B(r)/r} + r^2d\Omega^2,
\label{eq:MTmetric}
\end{equation}
an averaged null energy condition (ANEC) is defined by a line integral along a suitable radial null geodesic $I_{\rm ANEC} = \int (\rho+p_r)e^{-2\Phi}\,d\lambda = \int (\rho+p_r)e^{-\Phi}\,d\eta$ where $\lambda$ is an affine parameter and $\eta$ is the proper radial distance, defined by $d\eta^2 = (1-B(r)/r)^{-1}dr^2$. This line integral is insufficient to measure the total null energy contribution of the constituent matter in the spacetime. Visser, Kar and Dadhich (VKD) introduced a volume-integral formalism \cite{KDV2003}. For one asymptotic region, the spherical volume measure is $dV=4\pi r^2dr$. Since a wormhole contains two asymptotic regions, the corresponding volume integral is
\begin{equation}
\mathcal{I}_{\rm VKD} = 2\int_{r_0}^{\infty} 4\pi r^2(\rho+p_r)\,dr.
\label{eq:IKDVdefinition}
\end{equation}

Given a redshift function $\Phi(r)$ and shape function $B(r)$ as in Eq. (\ref{eq:MTmetric}), the stress energy-momentum components can be derived as
\begin{eqnarray}\label{eq:rhoMT}
&&\rho = \frac{B'}{8\pi r^2}~, p_r = \frac{1}{8\pi} \left[ -\frac{B}{r^3} + \frac{2}{r} \left(1-\frac{B}{r}\right) \Phi' \right],\\&&\nonumber
p_t = \frac{1}{8\pi} \left(1-\frac{B}{r}\right) \Big[ \Phi'' +\Phi' \Big( \Phi'+\frac1r \Big) \\&&
-\frac{B'r-B}{2r(r-B)} \Big( \Phi'+\frac1r \Big) \Big].
\end{eqnarray}

The wormhole throat is located at $r=r_0$, where $B(r_0)=r_0$. The flaring-out condition $B'(r_0)<1$ requires $\left (\rho + p_r\right)_{r=r_0}<0$. The integrand of Eq. (\ref{eq:IKDVdefinition}) can be derived from the field equations as
\begin{equation}
8\pi r^2(\rho+p_r) = (r-B) \frac{d}{dr} \ln\left[ \frac{e^{2\Phi}}{1-B/r} \right].
\label{eq:KDVtotalderivative}
\end{equation}

Substituting this in Eq. (\ref{eq:IKDVdefinition}) and integrating by parts one finds
\begin{eqnarray}
&&\mathcal{I}_{\rm VKD} = \left[(r-B) \ln\left(\frac{e^{2\Phi}}{1-B/r}\right) \right]_{r_0}^{\infty}
\\&&
- \int_{r_0}^{\infty}(1-B') \ln\left( \frac{e^{2\Phi}}{1-B/r} \right)dr .
\end{eqnarray}

At the throat, $r_0-B(r_0)=0$. If the redshift function has to be regular, the corresponding boundary contribution needs to vanish. The asymptotic boundary term also goes to zero for an asymptotically flat geometry, leaving
\begin{equation}
\mathcal{I}_{\rm VKD} = -\int_{r_0}^{\infty} (1-B') \ln\left[\frac{e^{2\Phi}}{1-B/r}\right]dr .
\label{eq:KDVtheorem}
\end{equation}

Visser, Kar and Dadhich considered a simple example, the Schwarzschild case with $B(r)=2M$, for which $\rho=0$ and the throat is located at $r_0=2M$. They also constructed geometries that differ from Schwarzschild only in a finite region $2M\leq r<a$, while matching smoothly to a Schwarzschild redshift function for $r \geq a$. They proved that the integrated NEC contribution can be made arbitrarily small as $a \rightarrow 2M^+$ each time, while retaining the required local NEC violation. Therefore, local exoticity around the throat and the total integrated violation of the energy condition, are categorically different.  \medskip

\paragraph*{\bf The Case of a Simpson-Visser Wormhole}
We first give an example and argue that not all wormholes fall in this category. We derive the volume integrated NEC for a Simpson-Visser metric defined as \cite{simpson}
\begin{eqnarray}\label{eq:SVoriginal}
&& ds^2 = -f(\ell)\,dt^2 + \frac{d\ell^2}{f(\ell)} + (\ell^2+a_{\rm SV}^2)d\Omega^2 ,\\&&
f(\ell) = 1-\frac{2M}{\sqrt{\ell^2+a_{\rm SV}^2}}.
\end{eqnarray}
We use the areal radius $r^2=\ell^2+a_{\rm SV}^2$ to rewrite Eq. \eqref{eq:SVoriginal} as
\begin{equation}
ds^2 = -\left(1-\frac{2M}{r}\right)dt^2 + \frac{dr^2}{\left(1-\frac{2M}{r}\right)\left(1-\frac{a_{\rm SV}^2}{r^2}\right)} + r^2d\Omega^2.
\label{eq:SVMTmetric}
\end{equation}
Comparing Eq. \eqref{eq:SVMTmetric} with the Morris-Thorne form we identify that
\begin{equation}
\Phi(r) = \frac12 \ln\left(1-\frac{2M}{r}\right), ~B_{\rm SV}(r) = 2M + \frac{a_{\rm SV}^2}{r} - \frac{2Ma_{\rm SV}^2}{r^2}.
\label{eq:SVshape}
\end{equation}

The associated throat is located at $\ell=0$ or $r_0=a_{\rm SV}$. At the throat, the flaring-out condition gives $\frac{4M}{a_{\rm SV}}-1<1$, which implies $a_{\rm SV}>2M$. Using the field equations, we derive the radial NEC as,
\begin{equation}
\rho_{\rm SV}+p_{r,\rm SV} = -\frac{a_{\rm SV}^2(r-2M)}{4\pi r^5}.
\label{eq:SVNEC}
\end{equation}

For the traversable branch $r \geq a_{\rm SV}>2M$, $\rho_{\rm SV}+p_{r,\rm SV} < 0$. We apply the VKD-integral formalism and evaluate, using Eq.~\eqref{eq:SVNEC},
\begin{align}\label{eq:SVvolume}
\mathcal{I}_{\rm VKD,~SV}
&=
8\pi
\int_{a_{\rm SV}}^\infty
r^2
\left[
-\frac{a_{\rm SV}^2(r-2M)}
{4\pi r^5}
\right]dr
\nonumber\\
&=
-2a_{\rm SV}^2
\left[
-\frac1r+\frac{M}{r^2}
\right]_{a_{\rm SV}}^\infty = 2(M-a_{\rm SV}).
\end{align}
(in the limit $\lim_{r\rightarrow\infty}$, $(-\frac1r+\frac{M}{r^2}) \to 0$). The near-horizon limit is found for $\lim_{a_{\rm SV}\rightarrow 2M^+}$, for which we find
\begin{equation}
\lim_{a_{\rm SV}\rightarrow2M^+}\mathcal{I}_{\rm SV} = -2M.
\label{eq:SVlimit}
\end{equation}

Thus, a Simpson-Visser wormhole cannot produce an arbitrarily-small volume integrated NEC measure. \medskip

\paragraph*{\bf Volume Integrated NEC and Reconstruction of the Redshift Function}
Motivated by this contrast, we explore whether one can prescribe the cumulative VKD integral as a function a priori and then use it to reconstruct the corresponding family of wormholes. For a generic Morris-Thorne structure, we denote the cumulative integral by $\Gamma(r)$,
\begin{equation}
\Gamma(r) \equiv -\int_{r_0}^{r} \left[1-B'(x)\right] \ln\left[\frac{e^{2\Phi(x)}}{1-B(x)/x}\right]dx.
\label{eq:Gamma_definition}
\end{equation}
\textit{Note:} $\Gamma(r)$ represents the VKD integral accumulated from the throat ($r = r_0$) up to an arbitrary radius $r$. $\Gamma(r_0) = 0$. Provided the integral converges, $\Gamma(\infty)=\mathcal{I}_{\rm VKD}$, i.e., the asymptotic value determines the total volume integrated NEC measure. The question is, whether one can have $\Gamma(\infty) = \delta_\infty^2 > 0$, while the NEC remains violated near the throat. We solve the integral equation above to write the redshift function as
\begin{equation}
\Phi(r) = \frac{1}{2} \ln\left(1-\frac{B(r)}{r}\right) - \frac{\Gamma'(r)}{2[1-B'(r)]}.
\label{eq:general_Phi_Gamma}
\end{equation}

\textit{Note : a positive value of $\Gamma(\infty)$ does not imply that the NEC is satisfied at all coordinate locations. It only means that the positive contribution accumulated away from the throat dominates.} For a regular traversable throat, $\Gamma(r)$ must remain finite at the throat, with $\Gamma(r_0)=0$, while its derivative must develop an integrable logarithmic behavior in order to cancel out the divergence of $\frac{1}{2}\ln[1-B(r)/r]$ in $\Phi(r)$. Therefore, it is easy to see that a constant $\Gamma'(r)$ case, although mathematically simple and tempting, is ruled out automatically.   \medskip

\paragraph*{\bf A Globally Convergent $\Gamma(r)$-Profile}
We now reconstruct an appropriate profile of $\Gamma$. From Eq. \eqref{eq:Gamma_definition},
\begin{equation}\label{gammaprime}
\Gamma'(r) = -\left[1-B'(r)\right]\ln\left[\frac{e^{2\Phi(r)}}{1-B(r)/r}\right].
\end{equation}
In order to maintain regularity near the throat, we Taylor-expand $1-B(r)/r$ and $e^{2\Phi(r)}$ at $r \to r_0^+$ as
\begin{eqnarray}
&& 1-\frac{B(r)}{r} = \frac{1-B'(r_0)}{r_0}(r-r_0) + O\!\left((r-r_0)^2\right), \\&&
e^{2\Phi(r)} = e^{2\Phi(r_0)}+O(r-r_0),~~ r\rightarrow r_0^+,
\end{eqnarray}
and simplify the RHS of Eq. (\ref{gammaprime}) as 
\begin{equation}
\frac{e^{2\Phi(r)}}{1-B(r)/r} = \frac{e^{2\Phi(r_0)}r_0}{\left[1-B'(r_0)\right](r-r_0)}\left[1+O(r-r_0)\right].
\end{equation}
Taking the logarithm, we find that
\begin{eqnarray}\nonumber
&& \ln\left[\frac{e^{2\Phi(r)}}{1-B(r)/r}\right] = -\ln(r-r_0) + \ln\left[\frac{e^{2\Phi(r_0)}r_0}{1-B'(r_0)} \right] \\&&
+ O(r-r_0) = -\ln(r-r_0) + O(1),
\end{eqnarray}
where $O(1)$ represents all finite term in the limit $r \rightarrow r_0^+$, in comparison to the divergent logarithmic first term. It follows that
\begin{equation}
\Gamma'(r) = \left[1-B'(r_0)\right]\ln(r-r_0)+O(1),~~ r\rightarrow r_0^+.
\end{equation}
Although $\Gamma'(r)$ has a (integrable) logarithmic behavior at the throat, $\Gamma(r)$ (and the cumulative NEC function) remains finite. Indeed, it is straightforward to check that
\begin{equation}
\int_{r_0}^{r}\ln(x-r_0)\,dx = (r-r_0)\ln(r-r_0)-(r-r_0),
\end{equation}
and therefore $\Gamma(r)-\Gamma(r_0) \to 0$ as $r \to r_0^+$. \medskip

We now choose a convenient profile for $\Gamma'(r)$ that preserves a logarithmic behavior at the throat and a convergence of the volume-integrated radial NEC contribution, as
\begin{equation}\label{gammaprime1}
\Gamma'(r) = -[1-B'(r_0)] \ln\left[1+\frac{L}{r-r_0} e^{-\frac{(r-r_0)}{L}} \right] + C e^{-\frac{(r-r_0)}{L}},
\end{equation}
where $C$ is a dimensionless parameter and $L > 0$ is a length scale. As $r \to r_0^+$, Eq. (\ref{gammaprime1}) gives
\begin{equation}
\Gamma'(r) = [1-B'(r_0)] \ln\left(\frac{r-r_0}{L}\right)+O(1),
\end{equation}
which is the logarithmic behavior required for a regularity at the throat. On the other hand, for $r \to \infty$,
\begin{equation}
\Gamma'(r) \sim \left[C-\frac{[1-B'(r_0)]L}{r-r_0}\right]e^{-\frac{(r-r_0)}{L}},
\end{equation}
so that the NEC integral converges. Introducing $u=(r-r_0)/L$, 
\begin{eqnarray}\nonumber
&&\Gamma(r) \equiv L\int_0^{u}\Big\{-[1-B'(r_0)] \ln\Big(1+\frac{e^{-u}}{u}\Big)+ C e^{-u}\Big\}du,\\&&\label{critical}
\Gamma(\infty) = L\left[C-[1-B'(r_0)]\mathcal{J} \right],\\&&
\mathcal{J}\equiv \int_0^\infty \ln\left(1+\frac{e^{-u}}{u}\right)du>0.
\end{eqnarray}
We note from Eq. (\ref{critical}) that the measure of a volume-integrated NEC can be shifted through zero, by simply varying $C$. Using Eq. (\ref{gammaprime1}) and Eq. (\ref{eq:general_Phi_Gamma}),
\begin{eqnarray}\nonumber
&& \Phi(r) = \frac12\ln\left(1-\frac{B(r)}{r}\right) + \frac{1-B'(r_0)}{2[1-B'(r)]} \ln\Big[1 +\frac{L}{r-r_0}\\&&\label{phirecon1}
e^{-\frac{(r-r_0)}{L}}\Big] - \frac{C e^{-\frac{(r-r_0)}{L}}}{2[1-B'(r)]}.
\end{eqnarray}
Near the throat, the logarithmic contribution in the second term cancels the corresponding divergence of the first term and keeps $\Phi(r_0)$ finite.  \medskip

As a simpler example, we consider a Morris-Thorne-like geometry and fix $B(r)=r_0$ and $B'(r)=0$, so that
\begin{equation}\label{mtrecon1}
\Gamma'_{\rm MT}(r) = -\ln\left[1+\frac{L}{r-r_0}e^{-\frac{(r-r_0)}{L}}\right] + C_{\rm MT}e^{-\frac{(r-r_0)}{L}},
\end{equation}
with $\Gamma_{\rm MT}(\infty) = L(C_{\rm MT}-\mathcal{J})$. The corresponding redshift function is
\begin{equation}\label{mtrecon2}
\Phi_{\rm MT}(r) = \frac12 \ln\left[\frac{r-r_0+L e^{-\frac{(r-r_0)}{L}}}{r}\right] - \frac{C_{\rm MT}}{2}e^{-\frac{(r-r_0)}{L}},
\end{equation}
which is finite at $r=r_0$. $\Gamma_{\rm MT}(\infty) > 0$ for all $C_{\rm MT}>\mathcal{J}$, explicitly showing the globally positive volume-integrated NEC category. \medskip

Therefore, one can construct a traversable wormhole starting from the globally volume integrated NEC, irrespective of whether the measure is positive or negative, preserving the NEC-violation near throat preserved. An arbitrarily-small NEC volume integral limit is recovered by taking $L\to 0^+$. For example, choosing $C_{\rm MT}=0$ gives $\Gamma_{\rm MT}(\infty)=-L\mathcal{J}<0$, and hence $\lim_{L\to 0^+}\Gamma_{\rm MT}(\infty)=0^-$. For every finite $L>0$, the redshift function remains finite at the throat, whereas for fixed $r>r_0$, $\Phi_{\rm MT}(r)\to \frac12\ln[(r-r_0)/r]$, recovering the Schwarzschild redshift function. \medskip

\paragraph*{\bf Classification : Global positivity versus local NEC violation}
The above construction clarifies the distinction between local NEC and volume-integrated NEC. The latter is determined by
\begin{equation}
\Gamma(\infty) = -\int_{r_0}^{\infty} \left[1-B'(r)\right] \ln\left[ \frac{e^{2\Phi(r)}}{1-B(r)/r} \right]dr.
\end{equation}
It is quite possible for the negative contribution near the throat to be compensated by positive contributions at larger radii, yielding $\Gamma(\infty) > 0$ despite obeying a flare-out condition $B'(r_0)<1 \rightarrow (\rho+p_r)_{r_0}<0$ near the throat. Based on this, we propose to classify traversable wormhole geometries according to the sign of the asymptotic value of $\Gamma$, as given in Table \ref{tab:KDV_classes}. For the convenient profile of $\Gamma(r)$ chosen through Eq. (\ref{gammaprime1}), the asymptotic value is derived as $\Gamma(\infty)=L\left[C-[1-B'(r_0)]\mathcal{J}\right]$, which gives the value of the global NEC volume integral of the wormhole geometry. It is additionally interesting to identify the constant $C$ as a critical parameter controlling the global energy characterization of the wormhole. Defining
\begin{equation}
C_{\rm crit}\equiv [1-B'(r_0)]\mathcal{J},
\end{equation}
we see that the three regimes $\Gamma(\infty)<0$, $\Gamma(\infty)=0$ and $\Gamma(\infty)>0$ correspond respectively to $C<C_{\rm crit}$, $C=C_{\rm crit}$, and $C>C_{\rm crit}$. Therefore, while the flare-out condition fixes the existence of local radial NEC violation at the throat, the volume-integrated radial NEC can indeed become negative, zero, or positive depending on the critical parameter $C$. \medskip

One can also keep the cumulative VKD function arbitrary, by taking
\begin{equation}
\Gamma'(r) = \left[1-B'(r_0)\right]\ln\left(\frac{r-r_0}{L}\right) + g(r).
\label{eq:free_Gamma_profile}
\end{equation}
$g(r)$ must be sufficiently well behaved at spatial infinity. The associated redshift function is
\begin{equation}
\Phi(r) = \frac12\ln\left(1-\frac{B(r)}{r}\right) - \frac{\left[1-B'(r_0)\right] \ln\left(\dfrac{r-r_0}{L}\right)+g(r)}{2[1-B'(r)]}.\label{eq:free_Gamma_Phi}
\end{equation}

\begin{table}[t]
\centering
\caption{Global classification based on $\Gamma(\infty)$.}
\label{tab:KDV_classes}
\begin{tabular}{c c c}
\hline
$\Gamma(\infty)$ & Local NEC & Volume-Integrated NEC \\
\hline
$<0$ & $< 0$ & $< 0$ \\
$\to 0^-$ & $\to 0^-$ & $< 0$ \\
$\to 0^+$ & $\to 0^-$ & $> 0$ \\
$>0$ & $< 0$ & $> 0$ \\
\hline
\end{tabular}
\end{table}

For convergence at spatial infinity, the large-$r$ behavior of $g(r)$ must also compensate the logarithmic growth of the first term in Eq. (\ref{eq:free_Gamma_profile}). Thus, asymptotically, Writing $A\equiv1-B'(r_0)$, we requir
\begin{equation}
g(r)=-A\ln\left(\frac{r-r_0}{L}\right)+\tilde g(r),~\int_{r_0}^{\infty}|\tilde g(r)|\,dr<\infty .
\end{equation}
The logarithmic terms then cancel asymptotically, while the remaining freedom in $\tilde g(r)$ controls the finite value of $\Gamma(\infty)$. This asymptotic condition alone does not determine the behavior of $g(r)$ at the throat; it must separately be chosen to remain finite there. A simple example is, once again,
\begin{equation}\label{grsolved}
g(r)=-A\ln\left[\frac{r-r_0}{L}+e^{-\frac{(r-r_0)}{L}}\right] + C\,e^{-\frac{(r-r_0)}{L}},
\end{equation}
for which $g(r)$ is finite at the throat and $\Gamma(\infty)$ is finite. Eq. (\ref{grsolved}) and (\ref{eq:free_Gamma_profile}) give
\begin{equation}
\Gamma'(r) = -A\ln\left[ 1+\frac{L}{r-r_0}e^{-\frac{(r-r_0)}{L}}\right] + C e^{-\frac{(r-r_0)}{L}},
\end{equation}
which is integrable at spatial infinity. Therefore, in this case $\Gamma(\infty)$ is finite and the parameter $C$ controls the finite volume-integrated NEC contribution without affecting throat regularity. \medskip

\paragraph*{\bf Smooth boundary matching}

The so-called free function $g(r)$ must also be constrained by the requirement of a smooth boundary matching \cite{darmois, israel}. For a generic Morris-Thorne type wormhole geometry with redshift function $\Phi(r)$ and shape function $B(r)$, we consider the reconstructed interior geometric patch to extend from the throat at $r=r_0$ to a matching surface at $r=a$. Beyond $r = a$, we take an exterior geometry for which the cumulative VKD integral vanishes, i.e., $\Gamma'_{\rm out}(r)=0$. We also define $Q(r) \equiv \frac{\Gamma'(r)}{1-B'(r)}$ and write Eq. (\ref{gammaprime}) as
\begin{equation}
\Phi(r) = \frac{1}{2} \ln\left(1-\frac{B(r)}{r}\right) -\frac{1}{2}Q(r).
\label{eq:Phi_Q_matching}
\end{equation}
Assuming that the spatial geometry is retained across the matching surface, the exterior redshift function corresponding to $\Gamma'_{\rm out}=0$ is
\begin{equation}
\Phi_{\rm out}(r) = \frac{1}{2}\ln\left(1-\frac{B(r)}{r}\right).
\end{equation}
Continuity of the induced metric at $r=a$ requires $\Phi_{\rm in}(a)=\Phi_{\rm out}(a)$, which, using Eq. \eqref{eq:Phi_Q_matching}, gives $Q(a)=0$. Equivalently, provided $1-B'(a)\neq0$, we have
\begin{equation}
\Gamma'(a)=0.
\label{eq:Gamma_prime_matching}
\end{equation}

To establish continuity of the extrinsic curvature at the matching hypersurface, we define the outward-pointing unit normal as $n^\mu = \left(0, \sqrt{1-\frac{B(r)}{r}},0,0\right)$. The $tt$ component of the extrinsic curvature in an orthonormal frame is $K_{tt} = -\sqrt{1-\frac{B(a)}{a}}\,\Phi'(a)$. Since we have assumed the spatial geometry to be preserved on both sides of the matching hypersurface, the angular components of $K_{ab}$ are automatically continuous. Therefore, any probable discontinuity in extrinsic curvature comes from $K_{tt}$ where $\Phi'$ is the contributing factor. Differentiating Eq. \eqref{eq:Phi_Q_matching}, we find that $\Phi'_{\rm in}(a)-\Phi'_{\rm out}(a) = -\frac{1}{2}Q'(a)$. Therefore, a continuity of $K_{tt}$ requires $Q'(a)=0$. In other words, for a smooth matching of the geometric patches, we need
\begin{equation}
Q(a)=0~;~Q'(a)=0 ~\Longrightarrow~ \Gamma'(a)=0~;~\Gamma''(a)=0.
\label{eq:Gamma_matching_conditions}
\end{equation}
The first condition $\Gamma'(a)=0$ ensures continuity of the induced metric and the second one $\Gamma''(a)=0$ eliminates a formation of thin-shell at the boundary. More importantly, these conditions give us constraints on the so-called arbitrary function $g(r)$. Using Eq. (\ref{eq:free_Gamma_profile}), we derive them as
\begin{equation}\label{eq:g_a_general}
g(a) = -[1-B'(r_0)]\ln\left(\frac{a-r_0}{L}\right),~g'(a) = -\frac{1-B'(r_0)}{a-r_0}.
\end{equation}
These requirements are independent of a particular choice of metric and they restrict the function only on the boundary hypersurface. For a spatial-Schwarzschild interior, $B(r)=r_0=2M$ and $B'(r_0)=0$, which leads to 
\begin{equation}
g(a) = -\ln\left(\frac{a-r_0}{L}\right),\qquad g'(a) = -\frac{1}{a-r_0}.
\label{eq:g_matching_SS}
\end{equation}

Similarly, for a Simpson-Visser-type shape function
\begin{equation}
B(r) = 2M+\frac{a_{\rm SV}^{\,2}}{r}-\frac{2Ma_{\rm SV}^{\,2}}{r^2},\qquad r_0=a_{\rm SV},
\end{equation}
we derive the requirements for smooth matching as
\begin{eqnarray}\label{eq:g_matching_SV}
&&g(a) = -\frac{2(a_{\rm SV}-2M)}{a_{\rm SV}}\ln\left(\frac{a-a_{\rm SV}}{L}\right),\\&&
g'(a) = -\frac{2(a_{\rm SV}-2M)}{a_{\rm SV}(a-a_{\rm SV})}.\label{eq:gprime_matching_SV}
\end{eqnarray}

\paragraph*{\bf Conclusion}
We have shown that the local profile of NEC near the throat and the global, volume integrated measure of NEC for a traversable wormhole need not share the same signature. Due to the flare-out condition, $(\rho+p_r)_{r_0}<0$, the radial NEC violation at the throat and ANEC violation along a null geodesic is unavoidable; but these do not dictate the sign of the corresponding integrated NEC. While it was already known that the integrated NEC can be made arbitrarily close to zero, we show that it can be positive as well. The central result is therefore

\medskip
\noindent\textit{The local exoticity at the throat does not determine the global exoticity or the signature of the volume integrated null energy measure of a traversable wormhole}
\medskip

We also show that by promoting the NEC volume integral to a cumulative function $\Gamma(r)$, it is possible to reconstruct a family of traversable wormholes from a prescribed global NEC profile. This reconstruction allows us to classify traversable wormholes categorically as $\Gamma(\infty) < 0$, $\Gamma(\infty) \to 0$ or $\Gamma(\infty)>0$. \medskip

\paragraph*{\bf Acknowledgements :} The author acknowledges Professor Sayan Kar for useful comments and suggestions. Acknowledgment is also given to the Vellore Institute of Technology for the financial support through its Seed Grant (No. SG20230027), 2023.

\end{document}